\documentclass[prd,twocolumn,showpacs]{revtex4}
\usepackage{amsfonts,amsmath,amssymb}

\newcommand{ \be}{\begin{equation}}
\newcommand{ \ee}{\end{equation}}
\newcommand{ \bea}{\begin{eqnarray}}
\newcommand{ \eea}{\end{eqnarray}}
\newcommand{ \eq}[1]{Eq.~(\ref{eq:#1})}
\newcommand{ \bm}   {\boldmath}
\newcommand{ \ubm}  {\unboldmath}

\newcommand{ \mysmall}[1]{\scriptscriptstyle #1} 

\newcommand{ \mw}{M_{\mysmall{W}}}
\newcommand{ \rw}{r_{\mysmall{W}}}
\newcommand{ \gw}{g_{\mysmall{W}}}
\newcommand{ \kw}{k_{\mysmall{W}}}
\newcommand{ \jw}{j_{\mysmall{W}}}
\newcommand{ \deltaw}{\delta_{\mysmall{W}}}

\begin{document}

\title{\bm $W$-propagator corrections to $\mu$ and $\tau$ leptonic decays\ubm}

\author{M.~Fael}
\email{matteo.fael@pd.infn.it}
\affiliation{Dipartimento di Fisica e Astronomia, Universit\`a di Padova}
\affiliation{Istituto Nazionale Fisica Nucleare, Sezione di Padova, I-35131 Padova, Italy}
\affiliation{Institut f\"{u}r Theoretische Physik, Universit\"{a}t Z\"{u}rich, CH-8057, Z\"{u}rich, Switzerland}

\author{L.~Mercolli}
\email{mercolli@astro.princeton.edu}
\affiliation{Princeton University, Department of Astrophysical Sciences, Princeton, NJ, 08544, USA}

\author{M.~Passera} 
\email{passera@pd.infn.it}
\affiliation{Istituto Nazionale Fisica Nucleare, Sezione di Padova, I-35131 Padova, Italy}

\begin{abstract} 
\noindent 
We derive the corrections induced by the $W$-boson propagator to the differential rates of the leptonic decay of a polarized muon and $\tau$ lepton. Results are presented both for decays inclusive of inner bremsstrahlung as well as for radiative ones, when a photon emitted in the decay process is measured. The numerical effect of these corrections is discussed. The definition of the Fermi constant $G_F$ is briefly reviewed. 
\end{abstract} 

\date{\today}
\maketitle

\section{Introduction}
\label{sec:intro}

The decay of the muon is one of the best known processes in particle physics. For more than 50 years it has played a fundamental role in our understanding of weak interactions, and increasingly precise measurements of its lifetime, combined with the calculation of radiative corrections, allowed to determine the Fermi constant $G_F$ with a relative precision of better than one part per million (ppm)~\cite{Webber:2010zf,Sirlin:2012mh,Marciano:1999ih}.

Muon and $\tau$ leptonic decays provide a clean probe to test the Standard Model (SM) and search for possible ``new physics" beyond it. 
Recent measurements of the muon decay spectrum reached the precision of $10^{-4}$~\cite{TWIST:2011aa}. The energy spectra of electrons and muons in leptonic $\tau$ decays have been measured to ${\cal O}(1\%)$~\cite{LEP} and significant improvements are expected at the upcoming SuperKEKB collider~\cite{Aushev:2010bq}. Progress is expected also in the analyses of muon and $\tau$ radiative leptonic decays, which provide a complementary tool to analyze the structure of the mediating current when inner bremsstrahlung is detected and measured. Precise radiative leptonic $\tau$ decay data may also allow to determine the $\tau$ dipole moments~\cite{Fael:2013ij}.

In this article we evaluate the corrections to muon and $\tau$ leptonic decays induced by the $W$-boson propagator. Calling $M$ and $m$ the masses of the initial and final charged leptons (neutrinos and antineutrinos are considered massless), we present both the leading contributions, of ${\cal O}(M^2/\mw^2)$, as well as the subleading ones, of ${\cal O}(m^2/\mw^2)$. 
They are of order
$m_{\tau}^2/\mw^2 \sim 5 \times 10^{-4}$,
$m_{\mu}^2/\mw^2 \sim 2 \times 10^{-6}$,
and
$m_e^2/\mw^2 \sim 4 \times 10^{-11}$.

We begin our analysis considering, in Sec.~\ref{sec:inclusive}, muon and $\tau$ leptonic decays when measurements are inclusive of inner bremsstrahlung and possible light-fermion pair production. We review the $W$-propagator corrections to integrated decay rates (also addressing a recent controversy in the literature) and present subleading, previously uncomputed, contributions to the energy-angle distribution of the final charged lepton. This section also contains a brief review of the definition of the Fermi constant $G_F$. In Sec.~\ref{sec:radiative} we derive the $W$-propagator corrections to the differential decay rates when a radiative photon is identified. Conclusions are drawn in Sec.~\ref{sec:conclusions}.

\section{Inclusive decays}
\label{sec:inclusive}

In this Section we consider the contribution of the $W$-boson propagator to the rates of muon and $\tau$ leptonic decays inclusive of the emission of radiation and possible light-fermion pair production accompanying the decay process. We denote them by 
\bea
\mu^- &\to& e^- \, \nu_{\mu} \, \bar{\nu}_e \, (\gamma),
\label{eq:muondecay}
\\
\tau^- &\to& l^- \, \nu_{\tau} \, \bar{\nu}_l \, (\gamma),
\label{eq:taudecay}
\eea
where $l=e$ or $\mu$. The theoretical prediction for these decay rates must include fully integrated inner bremsstrahlung and tiny contributions from the production of light-fermion pairs.

Let us focus our attention on muon decay. In the SM, the decay rate of~(\ref{eq:muondecay}) is~\cite{Ferroglia:2013dga}
\be
	\Gamma_{(\ref{eq:muondecay})} = \frac{G_{\mu}^2 M^5}{192 \pi^3} \, F \! \left( r^2 \right)
	\left( 1+ \delta_{\mu} \right)  \left[ 1+ \deltaw(M,m) \right] ,
\label{eq:muondecayformula}
\ee
where $r = m/M$, $\rw=M/\mw$,
\be
F(t) = 1 - 8t + 8t^3 - t^4 - 12t^2 \ln t  
\ee
is a phase-space factor and, as defined earlier, $M$ and $m$ are identified in muon decay with $m_{\mu}$ and $m_e$. Also, 
\be
¤\frac{G_{\mu}}{\sqrt 2} = \frac{g^2}{8 \mw^2} \left( 1+ \Delta r \right),
\ee
where $g$ is the SU$(2)_L$ gauge coupling constant and $\Delta r$ is the electroweak correction introduced by Sirlin in Ref.~\cite{Sirlin:1980nh}. The term $\delta_{\mu}$ is the QED correction evaluated in the Fermi $V$--$A$ theory; it includes the corrections of virtual and real photons up to ${\cal O} (\alpha^2)$, as well as the tiny contribution of the decay $\mu^- \to e^-  \nu_{\mu}  \bar{\nu}_e e^+ e^-$~\cite{Behrends:1955mb,Berman:1958ti,Kinoshita:1958ru,Roos:1971mj,QED2,vanRitbergen:1998hn,Ferroglia:1999tg}. Moreover,
\be
	\deltaw(M,m)  =  \frac{3}{5}  \, \rw^2  \frac{\left(1 - r^2 \right)^5}{F(r^2)} 
		+ \, {\cal O} \! \left( \rw^4 \right)
\label{eq:deltawdefinition}
\ee
is the tree-level correction induced by the $W$-boson propagator recently computed by Ferroglia, Greub, Sirlin and Zhang~\cite{Ferroglia:2013dga}. Its leading and next-to-leading contributions can be immediately derived from~\eq{deltawdefinition}: $(3/5)(M/\mw)^2$ and $(9/5)(m/\mw)^2$, respectively. While the leading one is well known in the literature~\cite{LeeYang,EcksteinPratt}, the next-to-leading term differs from that reported in earlier publications~\cite{Ho-KimPham, Krawczyk:2004na, Asner:2008nq}. We confirm the result in~\eq{deltawdefinition}, in agreement with Ref.~\cite{Ferroglia:2013dga}. 
We should add that while $(3/5)(m_{\mu}/\mw)^2  \sim 1.0 \times 10^{-6}$ is of the same magnitude as the present experimental relative uncertainty of the muon decay rate in \eq{muondecayformula}, 1.0 ppm, the subleading contribution $(9/5)(m_e/\mw)^2 \sim 7.3 \times 10^{-11}$ is out of experimental reach in the foreseeable future. Moreover, radiative corrections to~\eq{muondecayformula} of ${\cal O} (\alpha^3) \sim 10^{-7}$ and ${\cal O} (\alpha \, m_{\mu}^2/\mw^2) \sim 10^{-8}$ have not yet been computed.

The Fermi constant of weak interactions,  $G_F$, is defined from the muon lifetime $\tau_{\mu}$ evaluated in the Fermi $V$--$A$ theory 
\be
	{\cal L} = - \frac{G_F}{\sqrt{2}} \left[ \bar{\psi}_{\nu_{\mu}} \gamma^{\alpha} \left( 1- \gamma_5 \right) \psi_{\mu}
	\right] \left[ \bar{\psi}_e \gamma_{\alpha} \left( 1- \gamma_5 \right) \psi_{\nu_e} \right] + {\rm h.c.}
\ee
plus QED to leading order in the weak interaction coupling constant. We remind the reader that to leading order in $G_F$, but to all orders in $\alpha$, the radiative corrections to muon decay in the Fermi $V$--$A$ theory are finite after mass and charge renormalization~\cite{BermanSirlin1962}. Specifically, the present Particle Data Group (PDG) definition of $G_F$ is given by the relation~\cite{Beringer:1900zz,Sirlin:2012mh}
\be
\frac{1}{\tau_{\mu}} =  \frac{G_F^2 m_{\mu}^5}{192 \pi^3} \, F \! \left( \frac{m_e^2}{m_{\mu}^2} \right)
	\left( 1+ \delta_{\mu} \right).
\label{eq:GFdefinition}
\ee
This definition is independent of $\mw$, whereas earlier ones (see, for example, PDG 2010~\cite{Nakamura:2010zzi}) included the additional factor $[1+(3/5) m_{\mu}^2/\mw^2]$ on the r.h.s.\ of \eq{GFdefinition}. Since this factor does not arise in the Fermi theory framework, it is more natural not to include it in the definition in~\eq{GFdefinition}. Inserting the latest experimental value $\tau_{\mu} = 2 \, 196 \, 980.3(2.2)$ps~\cite{Webber:2010zf} and $\delta_{\mu} = -4.198 \, 18 \times 10^{-3}$~\cite{Sirlin:2012mh} into~\eq{GFdefinition}, one obtains $G_F = 1.166 \, 378 \, 7(6) \times 10^{-5} {\rm ~GeV}^{-2}$~\cite{Beringer:1900zz}. Also, identifying \eq{GFdefinition} with~\eq{muondecayformula} one finds the relation~\cite{Ferroglia:2013dga}
\be
	G_{\mu}^2 = G_F^2 / \left[ 1+ \deltaw(m_{\mu}, m_e) \right] ,
\label{eq:GFGmu}
\ee
with $\deltaw(m_{\mu}, m_e) = 1.04 \times 10^{-6}$ given by~\eq{deltawdefinition}.

The muon decay rate in~\eq{muondecayformula} can be immediately extended to the $\tau$ leptonic decays in~(\ref{eq:taudecay}) identifying $M$ with $m_{\tau}$ and $m$ with $m_e$ or $m_{\mu}$. The QED correction $\delta_{\mu}$ should also be replaced by $\delta_{\tau}$, the appropriate one for these decays,
while the electroweak corrections are the same as those contained in $G_{\mu}$ for muon decay~\cite{Marciano:1988vm}. Furthermore, in order to express these $\tau$ decay rates in terms of $G_F$, one should also replace $G_{\mu}$ in~\eq{muondecayformula} via~\eq{GFGmu}, thus obtaining
\be
	\Gamma_{(\ref{eq:taudecay})} = \frac{G_F^2 M^5}{192 \pi^3} \, F \! \left( r^2 \right)
	\left( 1+ \delta_{\tau} \right)  \left[ \frac{1+ \deltaw(M,m)}{1+ \deltaw(m_{\mu}, m_e)} \right].
\label{eq:taudecayformula}
\ee
Note that the leading contribution to $\deltaw(M,m)$, appearing in the numerator in square brackets, is independent of the flavor of the final lepton; it amounts to $(3/5)(m_{\tau}/\mw)^2  \sim 2.9 \times 10^{-4}$. The term $\deltaw(m_{\mu}, m_e)$ in the denominator, due to the relation between $G_{\mu}$ and $G_F$, has been kept for completeness, but it is of the same order of magnitude as the uncomputed radiative corrections of ${\cal O} (\alpha \, m_{\tau}^2/\mw^2) \sim 10^{-6}$. The hadronic corrections to~\eq{taudecayformula} are still missing too; they are of ${\cal O} (\alpha^2/\pi^2) \sim 10^{-5}$~\cite{vanRitbergen:1998hn,Actis:2010gg}.

The energy-angle distribution of the final charged lepton in the decays (\ref{eq:muondecay}) and (\ref{eq:taudecay}) of a polarized $\mu^-$ or $\tau^-$  at rest is~\cite{Kinoshita:1958ru}
\bea
	\frac{d^2 \Gamma_{(\ref{eq:muondecay},\ref{eq:taudecay})}}{ dx \, d\cos \theta_l} &=& 
				\frac{G_F^2 M^5}{192 \pi^3}  \frac{x \beta}{1+ \deltaw(m_{\mu}, m_e)} \times 
				\nonumber \\ 
				&& \!\!\!\!\!\!\!\!\!\!\!\!\!\!\!\! \left\{3x -2 x^2 +r^2 (3x -4) + f(x) \right.
				\nonumber \\
				&& \!\!\!\!\!\!\!\!\!\!\!\!\!\! \left. + \, \rw^2 \! \left[ 2x^2 -x^3 -2r^2  \! \left(1+x-x^2 +r^2 \right) \right]  \right.
				\nonumber \\
				&& \!\!\!\!\!\!\!\!\!\!\!\!\!\! \left. - \, \cos \theta_l \,\, x \beta  \left[ 2x-1 -3r^2  + g(x) \right. \right.
				\nonumber \\
				&& \!\!\!\!\!\!\!\!\!\!\!\!\!\! \left. \left. +\, \rw^2 \, x \left( x-2r^2 \right) \right] 
				+ {\cal O} \! \left( \rw^4 \right) \right\},
\label{eq:differentialnonradiative}
\eea
where $\beta \equiv  \lvert \vec{p}_l \rvert /E_l = \sqrt{1-4r^2/x^2}$, $p_l = (E_l, \vec{p}_l)$ is the four-momentum of the final charged lepton, $x=2E_l/M$ varies between $2r$ and $1+r^2$, $p$ and $n=(0,\hat{n})$ are the four-momentum and polarization vector of the initial muon or $\tau$, with $n^2=-1$ and $n \cdot p=0$, and $\cos \theta_l$ is the angle between $\hat{n}$ and $\vec{p}_l$. The corresponding formula for the decay of a polarized $\mu^+$ or $\tau^+$ is simply obtained inverting the sign in front of $\cos \theta_l$ in~\eq{differentialnonradiative}. The functions $f(x)$ and $g(x)$ are the QED radiative corrections; $f(x)$, contributing to the isotropic ($\theta_l$-independent) part, has been calculated up to ${\cal O} (\alpha^2)$, while $g(x)$, contributing to the anisotropic one, is known up to leading  ${\cal O} (\alpha^2)$ effects~\cite{Behrends:1955mb,Berman:1958ti,Kinoshita:1958ru,Roos:1971mj,Kinoshita:1957zz,QEDspectrum,Anastasiou:2005pn}. 
The hadronic corrections to~\eq{differentialnonradiative}, which are of  ${\cal O} (\alpha^2/\pi^2)$, were computed for the decay of the muon, but not yet for the $\tau$~\cite{Davydychev:2000ee}.
The terms proportional to $\rw^2$ are induced by the $W$-boson propagator. The leading ones, of ${\cal O} (\rw^2)$, agree with those of Ref.~\cite{Fischer:2002hn}; they are required by present studies of the Michel parameters in leptonic $\tau$ decays at Belle~\cite{Epifanov}. To our knowledge, the calculation of the subleading terms, of ${\cal O} (r^2 \rw^2)$, is new.

\section{Radiative decays}
\label{sec:radiative}

We now turn our attention to the contributions of the $W$-boson propagator to the decays
\bea
\mu^- &\to& e^- \, \nu_{\mu} \, \bar{\nu}_e \, \gamma,
\label{eq:muonraddecay}
\\
\tau^- &\to& l^- \, \nu_{\tau} \, \bar{\nu}_l \, \gamma,
\label{eq:tauraddecay}
\eea
when a radiative photon is detected and measured.
The tree-level SM prediction for the differential decay rates (\ref{eq:muonraddecay}, \ref{eq:tauraddecay}) of a polarized $\mu^-$ or $\tau^-$ at rest is
\begin{multline}
	\frac{d^6 \Gamma^0}{dx \, dy \, d\Omega_l\, d\Omega_\gamma}  =
	\frac{\,\alpha G_F^2 M^5} {(4 \pi)^6} \frac{x \beta}{1+ \deltaw(m_{\mu}, m_e)}  \bigg[ G_0(x,y,c) \\
	+ \,\, x \beta \, \hat{n} \cdot \hat{p}_l  \, J_0(x,y,c) \,\, + \,\, y \, \hat{n} \cdot \hat{p}_\gamma \, K_0(x,y,c) \bigg],
  \label{eq:radiativedecayrate}
\end{multline}
where $y = 2E_\gamma/M$, $E_{\gamma}$ is the energy of the photon, the final charged lepton and photon are emitted at solid angles $\Omega_l$ and $\Omega_{\gamma}$, respectively, with normalized three-momenta $\hat{p}_l$ and $\hat{p}_\gamma$, and  $c \equiv \cos \theta$ is the cosine of the angle between $\hat{p}_l$ and $\hat{p}_\gamma$. The corresponding formula for the radiative decay of a polarized $\mu^+$ or $\tau^+$ is simply obtained inverting the signs in front of the scalar products 
$\hat{n} \cdot \hat{p}_l$ and $\hat{n} \cdot \hat{p}_\gamma$ in~\eq{radiativedecayrate}.
The function $G_0$ and, analogously, $J_0$ and $K_0$, are given by
\be
  G_0 = \frac{4}{3 y z^2} \Big[ g_{0}(x,y,c) \,+\, \rw^2 \, \gw \! (x,y,c) + {\cal O} \! \left( \rw^4 \right) \Big],
  \label{eq:G0}
\ee
where
$
	z = xy \left( 1 - c \beta \right) \!/2.
$
The functions $g_{0}$, $j_{0}$, and $k_{0}$, computed in~\cite{KN1959radiative,Fronsdal:1959zzb,EcksteinPratt,Kuno:1999jp}, arise from the pure Fermi $V$--$A$ interaction, whereas $\gw$, $\jw$, and $\kw$ are the leading contributions of the $W$-boson propagator. Their explicit expressions are:
\begin{align}
	\gw & =
	z \big(-2 x^4 y-6 x^3 y^2+6 x^3 y z +4 x^3 y+2 x^3 z						\nonumber \\
	&-7 x^2 y^3+16 x^2 y^2 z+8 x^2 y^2-7  x^2 y z^2-2 x^2 y z 				\nonumber\\
	&-6 x^2 z^2 \! - \! 4 x^2 z-4 x y^4+14 x y^3 z+6 x y^3 -\!14 x y^2  z^2			\nonumber \\
	&-8 x y^2 z+4 x y z^3-12 x y z^2-8 x y z+6 x z^3+8 x z^2					\nonumber \\
	&-y^5+4 y^4 z +2 y^4-6 y^3 z^2-5 y^3 z+4 y^2 z^3 - \!4 y^2 z^2 				\nonumber \\
	&-4 y^2 z-y z^4+9 y z^3+14 y z^2-2 z^4-4 z^3 \big) 						\nonumber \\
	& + r^2 \big( 2 x^3 y^2+4 x^3 y z+6 x^2 y^3 +2 x^2 y^2 z-4 x^2 y^2			\nonumber \\
	&-8 x^2 y z^2-4 x^2 y z-4 x^2 z^2+6 x y^4-12 x y^3 z 						\nonumber \\
	& -8 x y^3-8 x y^2 z^2+4 x y^2 z+6 x y z^3-4 x y z^2-4 x y z 				\nonumber \\
	&+8 x z^3+4 x z^2+2 y^5-9 y^4 z-4 y^4+4 y^3 z^2 \! +12 y^3 z 				\nonumber \\
	&+5 y^2 z^3-4 y^2 z^2-2 y z^4+12 y z^3+4 y z^2-4 z^4 					\nonumber \\
	&-4 z^3+4 z^2 \big) 
	- 2 r^4\big( 2 x^2 y^2+4 x y^3-4 x y^2 z 								\nonumber \\
	&-2 x y^2+2 x y z+2 y^4-7 y^3 z-2 y^3+2 y^2 z^2 						\nonumber \\
	&+2 y^2 z-2 y^2-2 z^2 \big) 
	+ 4 r^6 y^2, 
	\label{eq:gw}
\end{align} 
\begin{align}
	\jw &= z \left( -4 x^3 y-10 x^2 y^2+10 x^2 y z +4 x^2 z-8 x y^3 \right. 			\nonumber \\
	&+21 x y^2 z-8 x y z^2+8 x y z-8 x  z^2-y^4+10 y^3 z 						\nonumber \\
	&\left. -y^3-11 y^2 z^2+2 y^2 z+2 y z^3-10 y z^2+4 z^3 \right) \! /2			\nonumber \\
	&+r^2\big( 4 x^2 y^2+8 x^2 y z+8 x y^3+4 x y^2 z-12 x y z^2				\nonumber \\
	&-8 x z^2 +4 y^4-5 y^3 z-8 y^2 z^2+4 y z^3-8 y z^2						\nonumber \\
	&+8 z^3\big) \!/2
	-4 r^4 y^2 \big(x+y-z \big), 
	\label{eq:jw}
\end{align} 
\begin{align}
	\kw &= z \left( -2 x^3 y+2 x^3 z-6 x^2 y^2+11 x^2 y z-6 x^2 z^2 \right.			\nonumber \\
	& -7 x y^3+18 x y^2 z+x y^2-17 x y z^2-2 x y z+6 x z^3					\nonumber \\
	&\left.-2 y^4+8 y^3 z-12 y^2 z^2+8 y z^3+2 y z-2 z^4 \right)  \! /2				\nonumber \\
	&+r^2 \big(4 x^2 y^2-4 x^2 z^2+8 x y^3-19 x y^2 z-8 x y z^2 				\nonumber \\
	&+8 x z^3+4 y^4-18 y^3 z+8 y^2 z^2+8 y^2 z+10 y z^3 					\nonumber \\
	&+6 y z^2-4 z^4\big)	\! /2											\nonumber \\
	&-2r^4 y \big( 2 x y-2 x z+2 y^2-7 y z+2 z^2 \big).
	\label{eq:kw}
\end{align} 
The kinematic limits for $x$, $c$, and $y$ are
\bea
	2r \leq & x & \leq 1+r^2, 					\nonumber \\
	-1 \leq & c & \leq 1,						 
\label{eq:fullrange}\\
	0  < & y & \leq y_\textup{max}(x,c),			\nonumber
\eea
where the maximum normalized photon energy is
\be
	y_{\max}(x, c) = \frac{2 \left(1+r^2-x \right)}{2-x + c  \, x \beta}.
\ee
However, every experimental setup has a minimum photon energy $E_{\gamma}^{\min} = y_{\min} (M/2)$ below which photons are not detected. As the constraint $y_{\min}<y_{\max}(x, c)$, necessary to measure radiative decays, leads to the bound $c < c_{\max}(x)$, with
\be
	c_{\max}(x)= \frac{2 \left(1+r^2-x \right) - \big(2-x\big) y_{\min}}{x \beta \, y_{\min}},
\label{eq:cmax}
\ee
the kinematic ranges of  $x$, $c$, and $y \! > \! y_{\min}$ are reduced to
\bea
	2r \leq & x & \leq 1+r^2, 						\nonumber \\
	-1 \leq & c & \leq \min \{1,c_{\max}(x)\},			
\label{eq:restrictedrange}\\
	 y_{\min} \leq & y & \leq y_{\max}(x,c).			\nonumber
\eea
We note that the terms in $G_0$, $J_0$, and $K_0$ proportional to $r^2$ cannot be neglected in the integrated decay rate. Indeed, the functions multiplying these $r^2$ terms generate a singular behavior in the $r \to 0$ limit after the integration over $c \equiv \cos \theta$: terms proportional to $r^2/z^2$ in $G_0$ (or $J_0$, $K_0$) lead to a nonvanishing contribution to the integrated decay rate since $\int \!  dc \, (1/z^2) \propto 1/z$ is evaluated at the integration limit $c \to 1$ where
$z \to xy \left( 1 - \beta \right) \!/2  \approx r^2 (y/x)$ for $x \gg 2r$.

One-loop radiative corrections to~\eq{radiativedecayrate} were computed in Refs.~\cite{Fischer:1994pn,Arbuzov:2004wr}.

If the initial $\mu^{\pm}$ or $\tau^{\pm}$ are not polarized, \eq{radiativedecayrate} simplifies to 
\be	
\frac{d^3 \Gamma^0}{dx \, dc \, dy}  =
	\frac{\,\alpha G_F^2 M^5} {(4 \pi)^6} \frac{8 \pi^2 \, x \beta}{1+ \deltaw(m_{\mu}, m_e)}  \,\, G_0(x,y,c).
\label{eq:radiativedecayrateunpolarized}
\ee
Integrating~\eq{radiativedecayrateunpolarized} over the kinematic ranges (\ref{eq:restrictedrange}) and dividing the result by the muon or $\tau$ total widths $\Gamma_{\mu,\tau}$ one obtains the branching ratios of the radiative decays~(\ref{eq:muonraddecay}, \ref{eq:tauraddecay}) for a given threshold $y_{\min}$. 
We note that these branching ratios contain mass singularities (and $\ln y_{\min}$)~\cite{EcksteinPratt,KN1959radiative}, but their presence does not contradict the Kinoshita--Lee--Nauenberg theorem, which applies only to total decay rates~\cite{Kinoshita:1958ru, KLN}.

In particular, for the branching ratio of radiative $\mu^+$ decays with a minimum detected photon energy $E_{\gamma}^{\min} = 10$~MeV we obtain $1.3 \times 10^{-2}$, to be compared with the experimental value $1.4(4) \times 10^{-2}$~\cite{Crittenden:1959hm}. A new preliminary measurement of this branching ratio has recently been reported by the PIBETA experiment~\cite{Pocanic}.
For radiative $\tau^-$ decays, with the same threshold $E_{\gamma}^{\min} = 10$~MeV, we obtain $1.84 \times 10^{-2}$ ($l = e$) and $3.67 \times 10^{-3}$ ($l = \mu$), to be compared with the values measured by the CLEO Collaboration,
$(1.75  \pm 0.06 \pm 0.17) \times 10^{-2}$ and 
$(3.61 \pm 0.16 \pm 0.35) \times 10^{-3}$, respectively, where the first error is statistical and the second one is systematic~\cite{Bergfeld:1999yh}.

\section{Conclusions}
\label{sec:conclusions}

We examined the corrections induced by the $W$-boson propagator to the rates of the leptonic decay of a polarized muon and $\tau$. We first analysed the experimental setup in which inner bremsstrahlung is not measured, reviewing the $W$-propagator correction to the integrated decay rates and the definition of the Fermi constant. Leading (${\cal O}(M^2/\mw^2)$) and subleading (${\cal O}(m^2/\mw^2)$) contributions to the energy-angle distribution of the final charged lepton were also presented ($M$ and $m$ are the masses of the initial and final charged leptons). The leading ones are already required by present studies of $\tau$ leptonic decays at Belle~\cite{Epifanov}. Comparisons with existing results were performed.

We then considered radiative decays, where the emission of a real photon is detected and measured, deriving the expression for the differential decay rate with respect to the angles and the energies of the final lepton and photon. The numerical impact of the $W$-propagator corrections to the differential rate of radiative muon decay is currently negligible because $m_{\mu}^2/\mw^2 \sim 10^{-6}$, whereas, at present, the relative error in the measurement of the electron spectrum is of ${\cal O}(1\%)$~\cite{Pocanic}. 
On the other hand, since in radiative $\tau$ decays $m_{\tau}^2/\mw^2 \sim 5 \times 10^{-4}$, $W$-propagator corrections are useful to control the present systematic uncertainty of the final-lepton spectra measured at Belle~\cite{EpifanovPrivate}, and their inclusion is expected to be important to interpret the precise results 
from the upcoming Belle-II experiment~\cite{Aushev:2010bq}. 

~\vspace{1mm}
\begin{acknowledgments}
We would like to thank S.~Eidelman, D.~Epifanov and A.~Sirlin for very useful comments.
The work of M.F.\ is supported in part by the Research Executive Agency of the European Union under the Grant Agreement  PITN-GA-2010-264564 (LHCPhenoNet).
L.M.\ is supported by a grant from the Swiss National Science Foundation. 
M.P.\ also thanks the Department of Physics and Astronomy of the University of Padova for its support. His work was supported in part by the Italian Ministero dell'Universit\`a e della Ricerca Scientifica under the program PRIN 2010-11, and by the European Programmes UNILHC (contract PITN-GA-2009-237920) and INVISIBLES (contract PITN-GA-2011-289442).
\end{acknowledgments}


\end{document}